\documentclass[twocolumn,aps,pra,superscriptaddress,nofootinbib]{revtex4-2}
\usepackage{bm,physics,amsmath,embedfile}

\usepackage[normalem]{ulem}

\usepackage[T1]{fontenc}
\usepackage{tikz}

\usepackage[utf8]{inputenc}
\usepackage{xcolor}
\usepackage{lipsum}
\usepackage{mathtools} 
\usepackage{amssymb} 
\usepackage{appendix}
\usepackage{braket}
\usepackage{graphicx}
\usepackage{hyperref}
\usepackage{multirow}
\usepackage{hhline}
\usepackage{ulem}

\begin{document}

\title{Two-dimensional bosonic droplets in a harmonic trap}

\author{Fabian Brauneis}
	\affiliation{Technische Universit\"{a}t Darmstadt$,$ Department of Physics$,$ 64289 Darmstadt$,$ Germany}

	\author{Artem G. Volosniev}
  \affiliation{Department of Physics and Astronomy$,$ Aarhus University$ ,$ Ny Munkegade 120$,$ DK-8000 Aarhus C$,$ Denmark}

  	\author{Hans-Werner Hammer}
	\affiliation{Technische Universit\"{a}t Darmstadt$,$ Department of Physics$,$ 64289 Darmstadt$,$ Germany}
	\affiliation{ExtreMe Matter Institute EMMI and Helmholtz Forschungsakademie
  Hessen f\"ur FAIR (HFHF)$,$ GSI Helmholtzzentrum f\"ur Schwerionenforschung GmbH$,$ 64291 Darmstadt$,$ Germany}

\begin{abstract}
    We investigate a system of bosons in a two-dimensional harmonic trap. In the limit of strong attractive interactions, the bosons make a droplet insensitive to external confinement. For weak interactions, in contrast, the ground state is given by the harmonic trap. In this work, we conduct a variational study of the transition between these two limits. We find that this transition occurs 
    at the critical interaction strength whose value is universal if scaled appropriately with the number of particles.  To connect the
    change in the properties of the system to the classical description of phase transitions, we analyze the static response of the Bose gas related to the isothermal compressibility. Finally, we perform numerically exact calculations for a few particles to demonstrate the effects of finite range interactions on this transition. We conclude that finite range effects wash out the point of transition.
\end{abstract}

\maketitle

\section{Introduction}

Twenty years ago, Hammer and Son predicted a universal many-body bound state for a large, but finite number of  attractively interacting bosons in two spatial dimensions (2D)~\cite{Hammer2004}. They found that upon addition of another boson to the system, the bound-state energy decreases by a factor of $E_{N+1}/E_{N}=8.567$ while the characteristic size shrinks by $R_{N+1}/R_{N}=0.3417$. The analysis was based on a mean-field approach that used the scale invariance of the contact interaction in 2D by introducing a renormalization group (RG) improved running coupling constant. These results were later confirmed by \textit{ab initio} calculations for particle numbers larger than twenty~\cite{Blume2005, Lee2006, Bazak_2018}. A possible platform for the experimental realization of such a system are cold atoms. Indeed, two-dimensional systems with bosons have been created experimentally, see e.g. Ref.~\cite{Hadzibabic2011}; few-body cold-atom systems have also been studied, albeit mainly  fermions~\cite{Bayha2020,Holten2021, Holten2022,brandstetter2023}. 

The crucial difference between the standard cold-atom experiments and the system of Ref.~\cite{Hammer2004} is the presence of a harmonic confinement, which introduces another length scale into the problem.
This means that the results of Ref.~\cite{Hammer2004} are only applicable for sufficiently strong interactions (corresponding to sufficiently small values of $R_{N}$) for which the effect of the trapping potential can be neglected. For weak interactions, the properties of the system are dictated by the trapping potential. The goal of our study is to provide insight into the transition between these two cases. In particular, we 
show that this transition contains information about universality of the many-body bound states in 2D,
providing a route for experimental studies of the corresponding physics.

To facilitate such studies, this paper provides an intuitive physical picture based upon a simple variational ansatz {(cf. Appendix \ref{App:Ansatz})} that incorporates the information about the trapping potential and the RG-improved coupling constant. {[An alternative ansatz that leads to similar conclusions is discussed in Appendix~\ref{App:OtherAnsatz}].} As consequence, both the harmonic oscillator ground state and the universal many-body bound state are accurately described in the limiting cases. The transition between the two states is induced by changing either of the two tunable experimental parameters: The scattering length (e.g. through Feshbach resonances~\cite{Chin2010}) or the trapping frequency, see Fig.~\ref{fig:Sketch}.  To characterize the transition, we introduce an observable which is closely related to the compressibility of the Bose gas~\cite{pines1994theory}. This observable, the static response, distinguishes the two states of the system and can be measured by changing the trapping frequency~\cite{Dalfovo1999, Pitaevskii2016}. Finally, we briefly discuss the impact of finite range effects on this transition by performing \textit{ab initio} calculations in the few-body sector. 

We note a relevant recent study~\cite{tononi2023gastosoliton} where bosonic droplets were studied on the surface of a sphere whose curvature naturally acts as an additional length scale. It was shown, in particular using Monte Carlo simulations, that for a finite but large particle number, an analogue of a first order phase transition occurs between a homogeneously distributed ground state and a localized droplet.
The results of Ref.~\cite{tononi2023gastosoliton} will be used to motivate certain aspects of the variational ansatz employed in the present paper.

The paper is structured as follows: In Sec.~\ref{sec:Formulation} we introduce the system and clarify the physics of the problem. Further, we explain the methods employed in the present study. Our main results are presented in Sec.~\ref{sec:Results}. Section~\ref{sec:Summary} briefly summarizes our findings and gives an outlook into future research perspectives. Further technical details are provided in {three} Appendices.

\section{Formulation and Methods}
\label{sec:Formulation}

\subsection{System}

We consider a system of attractively interacting bosons in a two-dimensional harmonic trap 
\begin{equation}
H=\int \mathrm{d} \Vec{x}\,\left[ \frac{\hbar^2}{2m}|\nabla \psi(\Vec{x})|^2+\frac{m\omega^2 \Vec{x}^2}{2} |\psi(\Vec{x})|^2+{U}\right],
    \label{eq:Ham}
\end{equation}
where $\psi(\Vec{x})$ is the bosonic field annihilation operator with $N=\int d^2x\, \psi^\dagger(\Vec{x})\psi(\Vec{x})$, the number of bosons. The operator ${U}$ is the boson-boson interaction ${U}=-\int d\Vec{x}'\psi^\dagger(\Vec{x})\psi^\dagger(\Vec{x}') V(\Vec{x}-\Vec{x}')\psi(\Vec{x}')\psi(\Vec{x})$, where the positive function $V$ describes the interaction potential. In the main part of this work, we assume zero-range interactions, $V(\Vec{x}-\Vec{x}')=g\delta(\Vec{x}-\Vec{x}')$ with $g>0$. Such interactions are typically used in the description of ultra cold gases~\cite{BRAATEN2006259, Bloch2008}. In Sec.~\ref{Subsec:FiniteRange}, we will also illustrate influence of finite range effects.

\begin{figure}
    \centering
    \includegraphics[width=\linewidth]{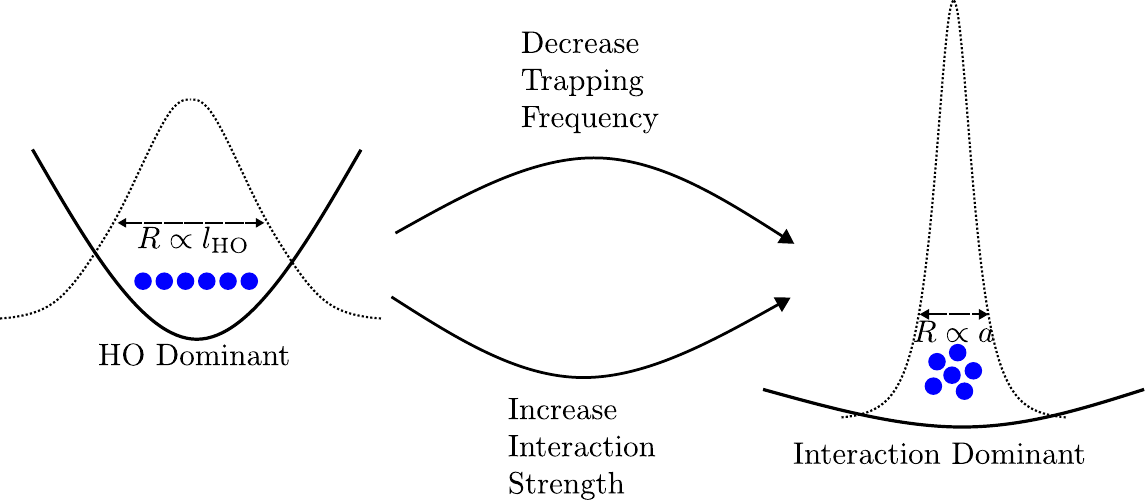}
    \caption{Sketch of the system. Blue spheres represent $N$ attractively interacting bosons. The solid curve is the trapping potential. The dashed curve illustrates the density of the bosons. For the {trap-dominated} regime, the density profile is given by a Gaussian. For the {interaction-dominated regime, the solution follows Ref.~\cite{Hammer2004} and has an exponentially decaying tail.} The characteristic length scale is given in the former case by the harmonic oscillator length $l_{\mathrm{HO}}$ while in the latter case it is proportional to the scattering length $a$. As sketched, a change from the {trap-dominated} to the {interaction-dominated} regime can be induced by either increasing the interaction strength or by decreasing the trapping frequency.}
    \label{fig:Sketch}
\end{figure}

We use a system of units such that $\hbar=m=1$. However, we keep the trapping frequency, i.e., we will give lengths in the units of the harmonic oscillator length {$l_{\mathrm{HO}}=1/\sqrt{\omega}$} and energies in the units of $\omega$. For a sketch of the system see Fig.~\ref{fig:Sketch}. 

The effective range expansion in two dimensions has a logarithmic dependence on the scattering length~\cite{Verhaar_1984}, or more precisely, on $\ln(ka)$ with $a$ the scattering length and $k$ the wave number. This implies that the interaction strength $g$ in two dimensions is a function of the considered length/momentum scales. In Ref.~\cite{Hammer2004} standard renormalization group arguments are used to introduce a coupling constant for a many-body problem which `runs' with the many-body ground state energy $E$ via 
\begin{equation}
    g(R)=-\frac{4\pi}{\ln(R^2 B_2)},
    \label{eq:grg}
\end{equation}
where $R$ is the characteristic length of the ground state, which is connected to $E$ by $E\sim 1/R^2$. Furthermore, $g(R)$ depends on the two-body binding energy energy in free space $B_2=-E_2$\footnote{Note that the two-body binding energy can be used to define the 2D scattering length {$a=2 e^{-\gamma}/\sqrt{B_2}$}~\cite{Verhaar_1984}, where $\gamma$ is the Euler-Mascheroni constant. {As the definition of the scattering length is ambiguous in two dimensions (cf. Ref.~\cite{Levinsen2015} for an alternative definition), we choose to always work with the two-body binding energy.}}. This equation implies that for a given two-body binding energy $B_2$, the size of the many-body bound state (its energy) determines the interaction strength such that the size is maximal (energy minimal). Therefore, in what follows we will use $B_2$ as the relevant scale for the interaction.

\vspace{1em}
\noindent\textit{Basic Physical Considerations}\vspace{1em}

\noindent We start by providing some physical intuition into the properties of the system. However, first let us summarize the results of Ref.~\cite{Hammer2004} where Eq.~(\ref{eq:Ham}) was studied without an external trapping potential, i.e., with $\omega=0$. In Ref.~\cite{Hammer2004}, first, the energy dependent coupling constant, Eq.~\eqref{eq:grg}, was introduced, which effectively replaced the bare interaction by the full multiple scattering series.
Then, the many-body problem was solved using a {renormalization-group-improved} mean-field ansatz, $\psi(\Vec{x})\sim \Psi(\Vec{x})$, where $\Psi(\Vec{x})$ a single particle orbital occupied by all particles:
\begin{equation}
    \label{eq:ansatz}\Psi(\Vec{x})=\frac{\sqrt{N}}{R_{\mathrm{free}}\sqrt{2\pi C}}f_{\mathrm{free}}(r/R_{\mathrm{free}})\,,
\end{equation}
with $r\equiv|\Vec{x}|$. The constant $C$ is determined by the particle number constraint, $C=\int d\rho \rho f_{\mathrm{free}}^2(\rho)$. 

The unknowns $R_{\mathrm{free}}$ and $f_{\mathrm{free}}$ in Eq.~(\ref{eq:ansatz})  are found by minimizing the expectation value of the Hamiltonian $H$
\begin{equation}
\label{eq:Energyfree}
    E(R_{\mathrm{free}})=\frac{A}{2C}\frac{N}{R_{\mathrm{free}}^2}-\frac{B}{4\pi C^2}\frac{g(R_{\mathrm{free}})N^2}{R_{\mathrm{free}}^2},
\end{equation}
where $A=\int d\rho \rho [f_{\mathrm{free}}'(\rho)]^2$, $B=\int d\rho \rho f_{\mathrm{free}}^4(\rho)$. 
The parameter $R_{\mathrm{free}}$ depends on $N$, and we shall use a subscript $N$ to indicate this fact explicitly when needed. {The two resulting equations ($dE/dR_{\mathrm{free}}=0$, $\delta E/\delta f_{\mathrm{free}}=0$) were solved separately, as the optimal value of $R_{\mathrm{free}}$ was found analytically from Eq.~(\ref{eq:Energyfree}). This expression allowed in turn to numerically determine the shape $f_{\mathrm{free}}$ independent of $R_{\mathrm{free}}$.}
For large values of $N$, this minimization demonstrated the existence of a universal bound state with the shape $f_{\mathrm{free}}$ independent on the number of particles. The corresponding energies and length scales obey $E_{N+1}/E_N=8.567$  and $R_{\mathrm{free}, N+1}/R_{\mathrm{free}, N}=0.3417$. 

Following the steps of Ref.~\cite{Hammer2004}, for a system with the harmonic trapping potential, we obtain instead of Eq.~(\ref{eq:Energyfree}) 
\begin{equation}
\label{eq:Energy}
    E(R)=\frac{A}{2C}\frac{N}{R^2}-\frac{B}{4\pi C^2}\frac{g(R)N^2}{R^2}
    +\frac{R^2}{l_{\mathrm{HO}}^4}\frac{ND}{2C}
\end{equation}
with $D=\int d\rho \rho^3 f^2(\rho)$. {Due to the additional scale $l_{\mathrm{HO}}$ from the trap, one cannot continue as in Ref.~\cite{Hammer2004}: A minimization of Eq.~\eqref{eq:Energy} under variation of $R$ and $f$ leads to two coupled equations which have to be solved simultaneously. Numerically, this is a nontrivial task. Nevertheless, we can use this equation to discuss} the physics of the system for limiting values of the new length scale $l_{\mathrm{HO}}$. For strong interactions, we have $R_{\mathrm{free}}\ll l_{\mathrm{HO}}$ and the effect of the harmonic trap is negligible. In this case, the system is described by Eq.~(\ref{eq:Energyfree}). For weak interactions, the size of the many-body bound state in free space becomes considerably larger than the harmonic oscillator length. Therefore, the state has to be restricted by the new length scale so that $R\approx l_{\mathrm{HO}}$ and $f\approx f_{\mathrm{HO}}$, the ground state of the 2D harmonic oscillator. In this case, the interaction term becomes negligible, and the ground state of the system is determined by the external trapping potential. We sketch these limiting cases in Fig.~\ref{fig:Sketch}.

\subsection{Methods}
\label{subsec:methods}

\noindent\textit{Variational Ansatz}

\vspace{1em}

\noindent Motivated by the discussion above, we use the variational ansatz
\begin{equation}
\label{eq:Ansatz}
    \Psi(\Vec{x})=\frac{1}{\mathcal{N}} \left[\alpha_{\mathrm{HO}}f_{\mathrm{HO}}(r)+\alpha_{\mathrm{free}}f_{\mathrm{free}}(r/R)\right]
\end{equation}
to study the transition between the two states of the system. The function $f_{\mathrm{HO}}$ describes the ground state of the harmonic oscillator,
\begin{equation}
    f_{\mathrm{HO}}(r)=
    \frac{1}{l_{\mathrm{HO}}\sqrt{\pi}}\,
    e^{-\frac{r^2}{2l_{\mathrm{HO}}^2}},
\end{equation}
and $f_{\mathrm{free}}$ is the shape found in Ref.~\cite{Hammer2004} for the universal many-body bound state; $\alpha_{\mathrm{HO}}$ and $\alpha_{\mathrm{free}}$ are variational parameters, $\mathcal{N}$ is a normalization coefficient, see App.~\ref{App:Ansatz}. The characteristic width of the state, is given by $R$. Following our physical insight we use $R=R_{\mathrm{free}}$ if $R_{\mathrm{free}}$ is smaller than the harmonic oscillator length. If it is larger, the characteristic length scale of the ground state is determined by the harmonic oscillator, so we use $R=l_{\mathrm{HO}}$ instead. We calculate $R_{\mathrm{free}}$ by minimizing the energy functional of Eq.~\eqref{eq:Energyfree} using the shape $f_{\mathrm{free}}$ for the calculation of $A,B,C$.

To find the values of  $\alpha_{\mathrm{HO}}$ and $\alpha_{\mathrm{free}}$ we minimize the expectation value of $H$ for a given interaction strength, i.e., for a given two-body binding energy in free space $B_2/\omega$. {By construction, this procedure reproduces the two limits: the trap-dominated physics (a non-interacting harmonic oscillator configuration) and the interaction-dominated physics, where the influence of the trap becomes negliglible. As Eq.~(\ref{eq:Ansatz}) has both of these limits, it} allows us to analyze the transition between the harmonic oscillator and the many-body bound state (see below). 
For a more detailed explanation of the employed variational ansatz including all relevant equations see Appendix~\ref{App:Ansatz}.

 {In addition, we also studied the system with an alternative ansatz discussed in Appendix \ref{App:OtherAnsatz} where both shapes scale with $R$, $\Psi(\Vec{r})\sim \alpha_{\mathrm{HO}}f_{\mathrm{HO}}(r/R)+\alpha_{\mathrm{free}}f_{\mathrm{free}}(r/R)$. This ansatz leads to properties of the transition that are consistent with those described in the main text.
}

\vspace{1em}

\noindent\textit{Configuration interaction method}

\vspace{1em}

\noindent In Sec.~\ref{Subsec:FiniteRange} we study the influence of finite-range effects on the transition. To that end, we perform an \textit{ab initio} calculation by employing the configuration interaction (CI) method~\cite{CremonPhDThesis, BjerlinPhdThesis}. Instead of the contact interaction, we use a Gaussian interaction potential, which reads as 
\begin{equation}
    V(\Vec{x})=\frac{g}{2\pi\sigma^2}e^{-\Vec{x}^2/2\sigma^2},
    \label{Eq:GaussianPot}
\end{equation}
$g>0$, $\sigma>0$, in Eq.~\eqref{eq:Ham}. To be able to compare with the results obtained with a contact interaction, we calculate $B_2$ by expanding the two-body Schr{\"o}dinger equation in free space using the harmonic oscillator eigenfunctions and subsequently diagonalizing the Hamiltonian matrix. We provide an explanation of our CI calculations in Appendix~\ref{App:CI}.

To estimate finite range effects of the Gaussian potential, we use the expression~\cite{Helfrich2011}:
\begin{equation}
\label{eq:EffectiveRange}
    B_2=-\frac{1}{r_{\mathrm{eff}}^2}W\left(-2 e^{-2\gamma}\frac{r_{\mathrm{eff}}^2}{a^2}\right),
\end{equation}
where $r_{\mathrm{eff}}$ is the effective range and $W$ is the product logarithm (also known as the Lambert $W$ function, defined as the solution of $z=we^w$ with $W(z)=w$). We estimate the scattering length using the semi-analytical expressions of Ref.~\cite{Jeszenszki2018}.  Then, we calculate the effective range by solving Eq.~\eqref{eq:EffectiveRange} using the scattering length and the numerically calculated two-body ground state energy in free space. 

\section{Results}
\label{sec:Results}

\subsection{Minimization of variational ansatz}
\label{subsec:A}

Here, we discuss the outcome of the minimization procedure for the variational ansatz discussed above.

\begin{figure}
    \centering
    \includegraphics[width=1\linewidth]{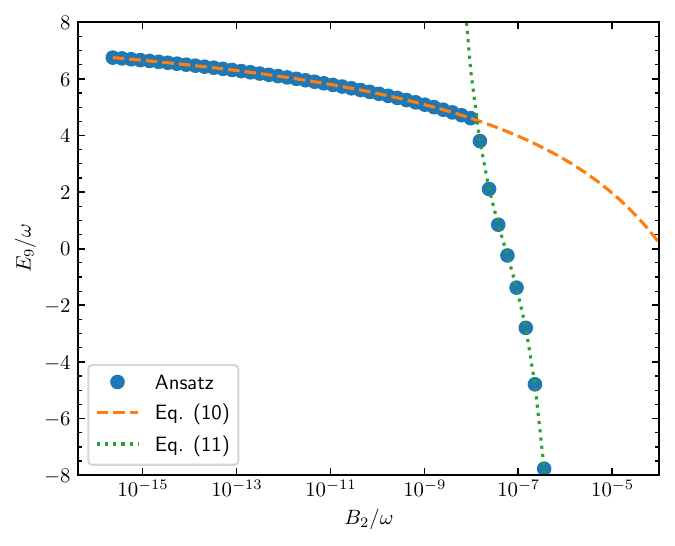}
    \caption{Energy for $N=9$ bosons calculated with our ansatz, Eq.~\eqref{eq:Ansatz}, together with the outcome of perturbation theory, see Eqs.~\eqref{eq:PerturbationHO} and~\eqref{eq:PerturbationUni}.}
    \label{fig:ComparisonPerturbationTheory}
\end{figure}

\vspace{1em}

\noindent\textit{Comparison with perturbation theory}

\vspace{1em}

\noindent In Fig.~\ref{fig:ComparisonPerturbationTheory} we show typical results from the minimization of our ansatz, Eq.~\eqref{eq:Ansatz}, together with perturbation theory for $N=9$ bosons as a function of the two-body energy in free space $B_2/\omega$\footnote{{Note that, for this number of bosons, the treatment of Ref.\cite{Hammer2004} deviates by only a few percent from the exact result~\cite{Bazak_2018}. This implies that $N=9$ is already large enough to study universal features of two-dimensional bosonic droplets.}}. Perturbation theory can be constructed for the two limiting cases discussed above, i.e., $B_2/\omega\to 0$ (weak boson-boson interactions) and $B_2/\omega\to\infty$ (strong boson-boson interaction).

For weak interactions, all bosons occupy the ground state of a harmonic oscillator and we treat the contact interaction as perturbation
\begin{equation}
\label{eq:PerturbationHO}
    E (B_2/\omega\to 0)=N\omega - \frac{g(R=l_{\mathrm{HO}})}{2}N^2\int d^2x f_{\mathrm{HO}}(\Vec{x})^4.
\end{equation}
Note that this integral can be evaluated analytically: $E(B_2/\omega\to 0)=N\omega - \frac{g(R=l_{\mathrm{HO})}}{4\pi}N^2$.
We have used that the characteristic width of the state of our system, $R$, is given by the harmonic oscillator length, i.e. $R=l_{\mathrm{HO}}$.
In the opposite limit of strong boson-boson attraction, we instead treat the harmonic oscillator potential as a perturbation to the universal many-body bound state
\begin{equation}
\label{eq:PerturbationUni}
    E(B_2/\omega\to \infty)=E(R_{\mathrm{free}})+\frac{\omega^2}{2}\frac{ND}{C}R_{\mathrm{free}}^2
\end{equation}
with $E(R_{\mathrm{free}})$ from Eq.~\eqref{eq:Energyfree}.
Figure~\ref{fig:ComparisonPerturbationTheory} demonstrates that perturbation theory is in excellent agreement with our ansatz. An inspection of the variationally optimized values of $\alpha_{\mathrm{HO}}$ and $\alpha_{\mathrm{free}}$ shows that for all particle numbers and all interactions one of these parameters is always zero, i.e., for every interaction strength the ground state of the system is described by either $f_{\text{HO}}$ or $f_{\text{free}}$ (see also Appendix~\ref{App:Ansatz}), which explains the excellent agreement between the variational results and perturbation theory in Fig.~\ref{fig:ComparisonPerturbationTheory}.
\\
\indent The sudden change of the variational parameters from $\{\alpha_{\mathrm{HO}}\neq0, \alpha_{\mathrm{free}}=0)\}$ to $\{\alpha_{\mathrm{HO}}=0, \; \alpha_{\mathrm{free}}\neq 0\}$ leads to a rather sharp transition between the two states for all particle numbers. This is a deliberately chosen feature of our ansatz motivated by the results of Ref.~\cite{tononi2023gastosoliton} for bosons on a sphere. In contrast, the alternative ansatz
discussed in Appendix~\ref{App:OtherAnsatz}, results in a smooth crossover for small particle numbers and becomes sharper as the particle number is increased\footnote{
It is beyond the scope of this work to provide a definitive answer as to which ansatz (the one in Eq.~(\ref{eq:Ansatz}) or the one in Eq.~(\ref{eq:App:2DBosons:AnsatzOther})) is better suited for our system. The results of the ansatze differ only in the vicinity of the transition between the trap-dominated and interaction-dominated regimes; beyond mean-field approaches are required for convincing results regarding the nature of this transition.
Following the conjecture of the  discontinuous transition~\cite{tononi2023gastosoliton} in a bubble trap we present only the results based upon the ansatz from Eq.~(\ref{eq:Ansatz}) in the main part of our manuscript.}
. 
This behavior is in agreement with the expectation that a sharp transition can occur only for large particle numbers, $N\gg1$; for small particle numbers the ground state energy should be a smooth function of the parameters\footnote{Here, we assume that there is no symmetry present that can lead to a crossing of energy levels.}.

We can use the perfect agreement of perturbation theory with our variational approach to study energies in a simple manner. In particular, we can use perturbation theory to estimate the transition point, i.e., the crossing in Fig.~\ref{fig:ComparisonPerturbationTheory}. Searching for the interaction strength at which the energies in Eqs.~(\ref{eq:PerturbationHO}) and~(\ref{eq:PerturbationUni}) are identical reveals that the transition happens at $R_{\mathrm{free}}=l_{\mathrm{HO}}$. This is also the outcome of the minimization calculation. Using Eq.~\eqref{eq:grg} and $g(R_{\mathrm{free}})=\frac{2\pi A C}{N B}$~\cite{Hammer2004} for large values of $N$, it follows that the transition should happen at\footnote{{Note that in Ref.~\cite{Hammer2004}, $R_{\mathrm{free}}$ could only be determined up to a constant $C_R$ of the order of one which leads to a $1/N$ correction in the transition point estimate.}} {
\begin{equation}
    \frac{\ln(B_2/\omega)}{N}=-2.148+\mathcal{O}(1/N).
    \label{eq:TransitionPoint}
\end{equation}
The fact that $\ln(B_2/\omega)/N$ becomes constant for large particle numbers} at the transition point suggests that this quantity is the proper measure of the interaction strength for our study. As we demonstrate below, the transition point indeed converges to this value as $N$ increases. This is also in agreement with Ref.~\cite{tononi2023gastosoliton}, which showed that with increasing particle number, the transition between the non-interacting state and the universal many-body bound state converges to $\ln(B_2/\Tilde{E})/N=-2.148+\mathcal{O}(1/N)$, where $\Tilde{E}$ is the relevant energy scale in that work. This showcases the universality of the transition point even for different systems.

\vspace{1em}
\noindent\textit{Ratio of the radii}
\vspace{1em}

\begin{figure}
    \centering
      \includegraphics[width=1\linewidth]{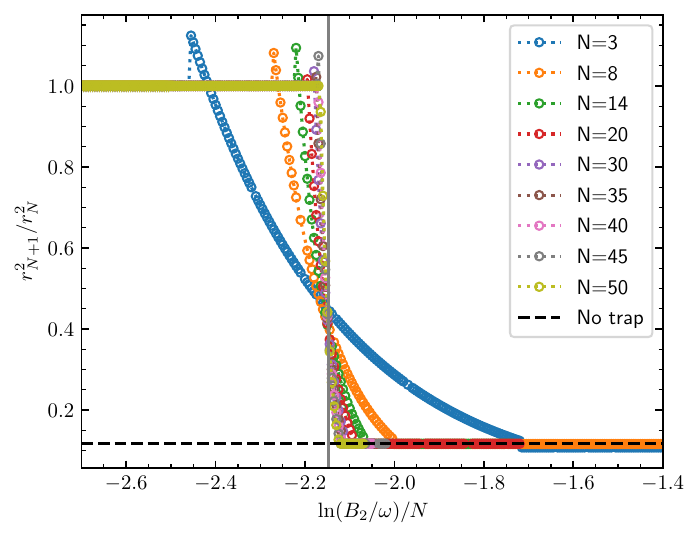}  
    \caption{Ratio of the mean-square radius $r^2_{N+1}/r^2_{N}$ as a function of the interaction strength given by $\ln(B_2/\omega)/N$. The symbols are the results obtained from the minimization of the energy with our ansatz, see Eq.~\eqref{eq:Ansatz}. The black dashed line is the universal prediction from Ref.~\cite{Hammer2004} {and the solid vertical line shows the transition point from Eq.~\eqref{eq:TransitionPoint}, $\ln(B_2/\omega)/N=-2.148$.} Dotted lines are added to guide the eye.}
    \label{fig:Radii}
\end{figure}

\noindent In Fig.~\ref{fig:Radii} we show the ratio of the mean-square radius $r^2_{N+1}/r^2_{N}$ with\footnote{Note that $\Psi(\Vec{x})$ depends on $N$, see Eq.~\eqref{eq:ansatz}.} 
\begin{equation}
    r^2_{N}=\frac{1}{N}\int d^2x\, \Psi^2(\Vec{x})\Vec{x}^2,
\end{equation}
which is one of the hallmark measures of the bound-state universality in 2D.
For weak interactions, the ratio corresponds to non-interacting bosons in a harmonic oscillator ($r^2_{N+1}/r^2_{N}=1$) while for strong interactions we recover the universal ratio predicted in Ref.~\cite{Hammer2004} ($r^2_{N+1}/r^2_{N}=0.116$). In between these limits, we see that the ratio first increases to values slightly higher than unity and then decreases rapidly. 
The sudden increase is an artifact of our ansatz caused by the transition from the trap-dominant to the interaction-dominant states, which happens for larger particle numbers for weaker interactions (as can be seen from the figure). It can be directly understood from the minimization procedure: As soon as the two-body binding energy is large enough such that $R_{\mathrm{free}}\leq l_{\mathrm{HO}}$, we have a state dominated by the interactions. Now, we can calculate the mean-square radius for $R_{\mathrm{free}}=l_{\mathrm{HO}}$
\begin{equation}
    r_N^2=2\pi l_{\mathrm{HO}}^2\int d\rho \rho^3 f_{\mathrm{free}}^2(\rho)\approx 1.2l_{\mathrm{HO}}^2\,,
\end{equation}
i.e., the maximum value of $r^2_{N+1}/r^2_{N}$ we observe. 
The decrease of the ratio that follows happens when the $N+1$ system is dominated by the interactions, while the $N$ system is still in an oscillator state.

Further, Fig.~\ref{fig:Radii} confirms our estimate for the transition point from perturbation theory, Eq.~\eqref{eq:TransitionPoint}. Indeed, according to these data, the transition occurs approximately at $\ln(B_2/\omega)/N\approx-2.2$ for all particle numbers. A weak dependence on $N$ is visible only for small systems; for large values of $N$, the transition point converges towards the value expected from Eq.~\eqref{eq:TransitionPoint}. 

Due to the logarithmic scaling with $B_2/\omega$, very large values of the scattering length $a\propto B_2^{-1/2}$ with a high precision are needed to detect the transition. Here, the harmonic trapping confinement can help to alleviate this issue: One can tune the value $B_2/\omega$ also by changing the trapping frequency.
This additional degree of freedom might facilitate experimental studies of this transition.
{
Note that such studies can provide information about the transition point even when experiments are performed in quasi-2D. Indeed, as long as the interaction strength is in the transition region, we have $R\approx l_{\mathrm{HO}}$, and the system will not be affected by additional 3D effects. This contrasts with the validation of the free-space results of Ref.~\cite{Hammer2004}, where one must ensure that the characteristic sizes of the bound states can be measured -- a non-trivial task due to the exponential sensitivity of the radii to the binding energy.}

\subsection{Transition indicator: static response}

Even with the help of tuning both, the trapping frequency and the scattering length, detecting the transition through the ratio of the mean-square radii (or also the energies) might be very challenging. One must not only maintain high level of control over the experimental parameters (either $B_2$ or $\omega$), but do so for two experimental runs with different particle numbers. In this subsection, we note that the information about the transition between the two states could actually be extracted in experiments that operate with a single value of $N$. We demonstrate this fact by introducing an observable motivated by the standard description of phase transitions. It is also an illustration of a general observation that concepts developed for the description of phases in the thermodynamic limit could be adapted for distinguishing the two states of our system, even though it is finite. 

Here, we use the compressibility, which is one of the standard markers of phase transitions~\cite{schwabl2006statistical}. Using linear response theory, one can show that the compressibility of a (uniform) system is strongly related to the so-called static response (or polarizability) $\chi$ which describes the response of a system's density to a static force field\footnote{This is known under the name of the compressibility sum rule.}~\cite{pines1994theory}. For a harmonically trapped system, we can directly calculate $\chi$ from the change of the mean-square radius upon varying the trapping frequency (under these circumstances $\chi$ is also called monopole compressibility)~\cite{Dalfovo1999, Pitaevskii2016}:\footnote{In practice, it might be more convenient to perform calculations using a system of units in which $\omega=1$. To calculate $\chi$ in this case,  one should express the derivative with respect to the trapping frequency in terms of a derivative with respect to $B_2$, the interaction strength parameter. This transformation leads to the following definition of $\chi$: $
    \chi=N\left(B_2\frac{\partial \braket{r^2}}{\partial B_2}+\braket{r^2}\right).$}
\begin{equation}
    \chi=-2N\frac{\partial \braket{r^2}}{\partial (\omega^2)}.
    \label{eq:Compress}
\end{equation}
This quantity allow us to distinguish the two different phases without relying on the ratios of energies or radii, which can certainly make a detection of the transition point significantly easier.

\begin{figure}
    \centering
    \includegraphics[width=1\linewidth]{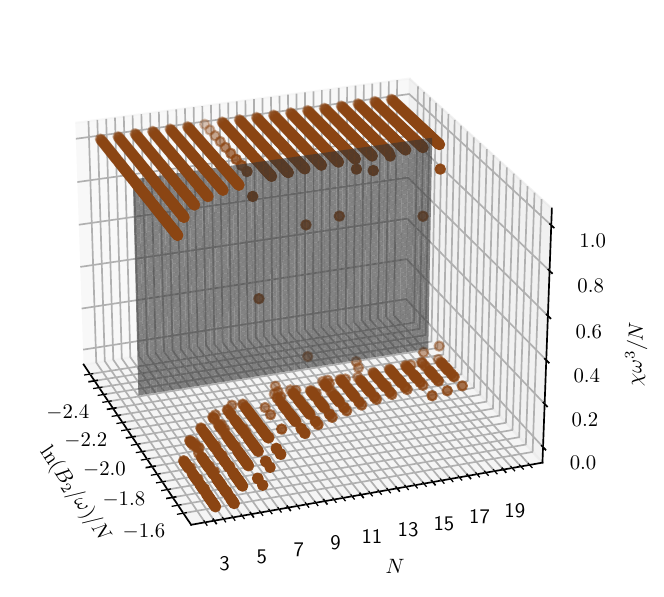}
    \caption{Static response per particle, $\chi\omega^3/N$, see Eq.~\eqref{eq:Compress}, as a function of the interaction strength $\ln(B_2/\omega)/N$ and particle numbers $N$. All points are calculated with our variational ansatz, Eq.~\eqref{eq:Ansatz}. {The transparent vertical wall corresponds to the transition point from Eq.~\eqref{eq:TransitionPoint}, $\ln(B_2/\omega)/N=-2.148$.} Note that the calculation of the static response close to the transition point requires a high numerical precision for the evaluation of the derivative, see Eq.~\eqref{eq:Compress}.} 
    \label{fig:Compressibility}
\end{figure}

For weak interactions, our system is described by the harmonic oscillator ground state whose mean-square radius naturally follows the trapping frequency. Therefore, we expect the static response to be $\chi\omega^3=N$. In the opposite limit of strong interactions, we can use the result of Ref.~\cite{Hammer2004} and approximate the behavior $\braket{r^2}\propto \frac{1}{B_2}$, i.e., there is no influence of the trapping frequency on the state and hence $\chi=0$. 
Therefore, the static response does indeed show a very different behavior for the two states of our system and can be used as an indicator for the transition.

We show a plot of the static response obtained from the minimization procedure in Fig.~\ref{fig:Compressibility}. We see exactly the behavior discussed above: For weak interactions, the static response remains constant at $\chi\omega^3/N=1$ up until the transition point. After that, the static response is zero (within our numerical accuracy). Therefore, the observable $\chi$ allows one to see that a) there is a transition between two different states of the system with very different properties, b) that the interaction strength needed for the transition converges towards $\ln(B_2/\omega)/N\approx-2.15$ with increasing particle numbers, in agreement with our estimate from perturbation theory, Eq.~\eqref{eq:TransitionPoint}.

\subsection{Influence of finite range effects}
\label{Subsec:FiniteRange}

Even though zero-range potentials describe cold-atom systems faithfully~\cite{BRAATEN2006259, Bloch2008}, it is known that in two-dimensional systems already small effective range corrections can have a large impact on the properties of a few-body system. For an example see, e.g., Ref.~\cite{Helfrich2011}, which considers a three-boson system in free space. Therefore, one should study the influence of finite range effects on the transition discussed above\footnote{Note that for large finite range effects, the many-body bound state is no longer universal, see Ref.~\cite{Hammer2004}}. To motivate such studies, here, we employ a Gaussian potential of finite width, Eq.~\eqref{Eq:GaussianPot}, and perform \textit{ab initio} calculations with the CI method, see Sec.~\ref{subsec:methods}.

We use widths of the Gaussian potential in the order of the harmonic oscillator length, $\sigma=l_{\mathrm{HO}}/2$ and $\sigma=l_{\mathrm{HO}}$. These values cause significant effective range corrections as we will show below. At the same time, the width of the potential acts as a momentum space cutoff, enabling \textit{ab initio} calculations with the CI method within the transition region. Note that for strong attractive interactions, such calculations become computationally very expensive. Therefore, it is challenging to explore the (quasi-~)zero-range limit with Gaussian interactions\footnote{In practice, weak attractive interactions can pose challenges as well, since we calculate the two-body binding energy numerically. For weak interactions, this energy becomes exponentially small~\cite{Simon1976}, causing loss of accuracy in numerical calculations.}. {While it is clear that finite range interactions break universality in two dimensions, it is \textit{a priori} unclear how this modifies the transition between the trap dominated and the interaction dominate regime. It appears interesting to study how the universal transition point is changed. By comparing the two widths, we can discuss the finite range effects on the transition on a qualitative level.}

\begin{figure}
    \centering
    \begin{minipage}{\linewidth}
    \centering
    \includegraphics[width=1\linewidth]{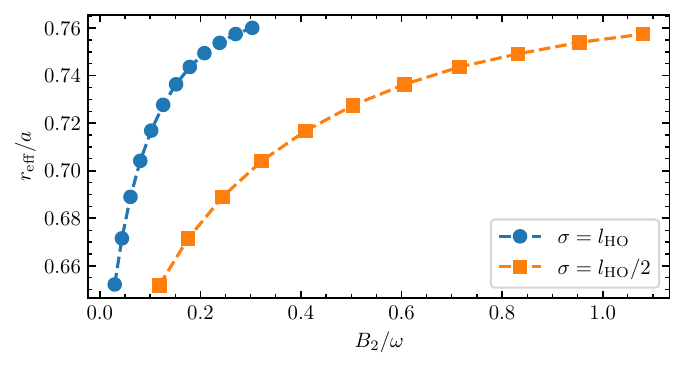}
    \end{minipage}
    \begin{minipage}{\linewidth}
    \centering
    \includegraphics[width=1\linewidth]{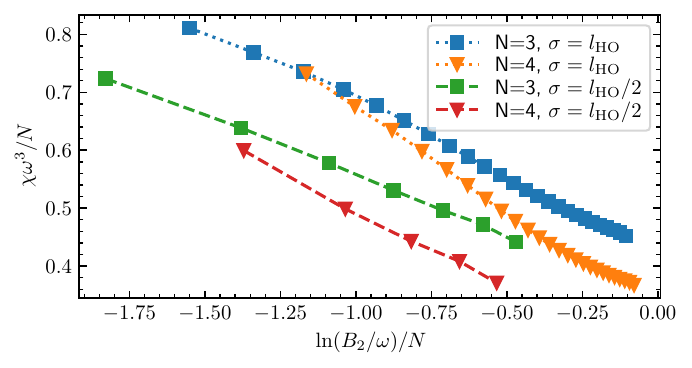}
    \end{minipage}
    \caption{Upper panel: Ratio of the effective range to the scattering length as a function of $B_2/\omega$ for a Gaussian potential with $\sigma=l_{\mathrm{HO}}/2$ and $\sigma=l_{\mathrm{HO}}$. 
    Lower panel: Static response per particle, Eq.~\eqref{eq:Compress}, calculated with a CI method as a function of the interaction strength given by $\ln(B_2/\omega)/N$ (for details on the method see App.~\ref{App:CI}). For the interaction we use a Gaussian potential, Eq.~\eqref{Eq:GaussianPot}. Squares show results for $\sigma=l_{\mathrm{HO}}$ while triangles are for $\sigma=l_{\mathrm{HO}}/2$. The dashed (dotted) lines are added to guide the eye.}
    \label{fig:CICompressibility}
\end{figure}

Let us first estimate the size of the finite range corrections. We show in Fig.~\ref{fig:CICompressibility} the ratio of the effective range and the scattering length as a function of $B_2/\omega$\footnote{Note that the oscillator units enter the two-body problem via the width of the Gaussian potential which we define with respect to $l_{\mathrm{HO}}$.}. As one can clearly see, the role of the effective range can be very large for these parameters. Indeed, for all considered values of $B_2/\omega$ the influence of the effective range cannot be neglected. Furthermore, Eq.~\eqref{eq:EffectiveRange} only considers finite range corrections up to the effective range. 
Since the values of the effective range are so large, higher-order terms, such as the shape parameter, should also be significant. However, for our study, the calculation of such terms is not important as we are only interested in comparing the size of the finite range effects for different values of $\sigma$. For the smaller value of $\sigma$ one can see that the effective range for the same value of $B_2/\omega$ is smaller, implying that finite range effects are of lesser relevance.

In Fig.~\ref{fig:CICompressibility} we present the static response for $N=3, 4$ particles for the two different widths of the Gaussian potential as a function of $\ln(B_2/\omega)/N$. We can see that the static response does decrease with increasing binding energy. Since we now perform an \textit{ab initio} calculation with a finite-range potential, we do not see a sharp transition for such small particle numbers as in the previous section. Instead, we see a faster decay of the static response for larger particle numbers. This observation is in agreement with  Ref.~\cite{tononi2023gastosoliton} where a sharper transition between a localized droplet state and a homogeneously distributed state was observed when the particle number was increased. 

The effects of the  finite range are visible in Fig.~\ref{fig:CICompressibility} when comparing the results for the two values of $\sigma$. For the largest $\sigma$, the decrease of the static response is pushed towards larger values of the binding energy. By comparing the values of  $\ln(B_2/\omega)/N$ in Fig.~\ref{fig:CICompressibility} with the critical interaction strength $\ln(B_2/\omega)/N\approx-2.15$, we conclude that significantly stronger attraction is needed to affect the static response for a system with finite-range interactions. Further \textit{ab initio} calculations with larger particle numbers and smaller finite ranges are needed to recover the transition behavior described in Sec.~\ref{subsec:A}. Such studies will further assess the impact of finite range corrections on the physics of two-dimensional droplets.

\section{Summary and Outlook}
\label{sec:Summary}

In this work we studied the transition of attractively interacting bosons in a two-dimensional harmonic oscillator from a trap-dominated state (the ground state of a harmonic oscillator) for weak interactions to a universal many-body bound state for strong interactions. With a physically motivated variational ansatz, we showed that this transition happens at $\ln(B_2/\omega)/N\simeq -2.15$ for large particle numbers. We argued that the transition can be induced by tuning two
parameters: The scattering length and the trapping frequency. Further, the classical description of phase transitions motivated us to study the static response -- an observable closely related to isothermal compressibility. We showed that it varies significantly between the two states and argued that it can be used as an indicator of the transition. Finally, we discussed the influence of finite range effects in few-body \textit{ab initio} calculations with the configuration interaction method. We found that large finite range effects modify the point of transition. 

Further \textit{ab initio} calculations with finite range interactions are needed to better understand their influence on the transition, in particular since Ref.~\cite{Helfrich2011} showcases strong influence of finite range effects for the three-body system in free space. Such calculations might also be important for corresponding experimental studies. Therefore, more sophisticated truncation schemes such as the importance-truncation CI should be employed (see e.g. Refs.~\cite{Roth2009, chergui2023superfluid, Bengtsson2020} for applications in the context of ultracold atoms). Furthermore, techniques such as renormalized interactions can be useful to improve the convergence. {
These approaches would enable the treatment of stronger interactions and/or narrower Gaussian potentials~\cite{Christensson2009, brauneis2024comparisonrenormalizedinteractionsusing}, allowing for a more accurate description of the finite-range system.} Alternatively, Monte-Carlo calculations with finite-range interactions could be used for the study.

Another interesting feature is the increase in sharpness of the transition. 
Indeed, our results predict that a rapid transition between the trap dominated and interaction dominated regimes  at $\ln(B_2/\omega)=-2.148$ as $N$ is increased. It would be worthwhile establishing when the transition of the considered few-body systems can be treated as a genuine phase transition. Then one could determine its order and study the dependence on external parameters such as the trapping potential. 
Such a study would help determine which of the two ansatze (see Eqs.~(\ref{eq:Ansatz}) and~(\ref{eq:App:2DBosons:AnsatzOther})) more accurately describes the physics of two-dimensional bosons in a trap\footnote{
For example, the static response with the ansatz in Eq.~(\ref{eq:Ansatz}) (cf. Fig.~\ref{fig:CICompressibility}),
features sharp transition already for small 
$N$, consistent with Ref.~\cite{tononi2023gastosoliton}. The ansatz from App.~\ref{App:Ansatz} leads to a smooth transition between the trap-dominated state and the interaction-dominated
one at small particle numbers $N$, which becomes sharper as $N$ is increased.}
.
Furthermore, Ref.~\cite{tononi2023gastosoliton} and our work focused on the ground state of the system. However, excited states might also provide valuable insight into the (phase) transition like in Ref.~\cite{Bayha2020, Bjerlin2016} where excited states for a finite system were interpreted as precursors of a Higgs mode. We leave such studies to future work.

\vspace{1em} 

\section*{Acknowledgements}
We thank Stephanie Reimann and the Lund cold atom group for giving us access to their configuration interaction code. 
We thank Micha{\l} Suchorowski for useful discussions.
H.W.H. was supported in part by Deutsche Forschungsgemeinschaft (DFG, German Research Foundation) - Project-ID 279384907 - SFB 1245 and by the German Federal Ministry of Education and Research (BMBF) (Grants No. 05P21RDFNB and 05P24RDB).

\newpage

\appendix

\widetext
\section{Details on the variational ansatz}
\label{App:Ansatz}

\begin{figure}
    \centering
    \begin{minipage}{0.5\linewidth}
    \centering
    \includegraphics[width=1\linewidth]{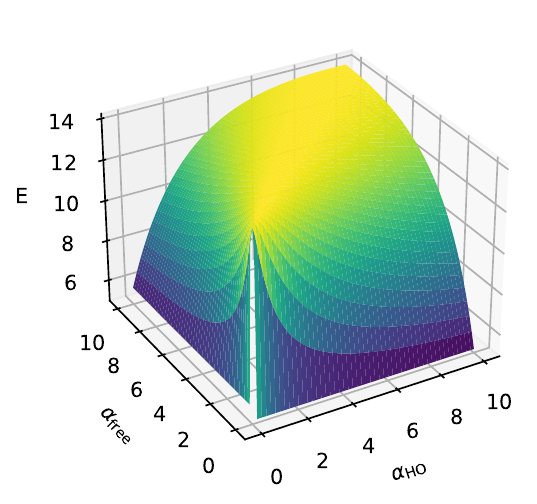}
    \end{minipage}%
    \begin{minipage}{0.5\linewidth}
    \centering
    \includegraphics[width=1\linewidth]{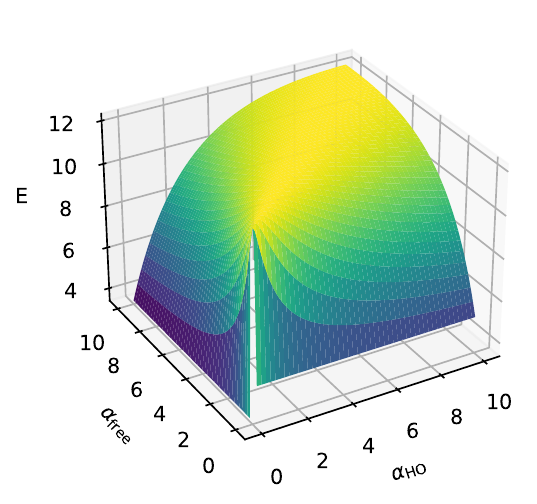}
    \end{minipage}
    \caption{Energy calculated with our ansatz, Eq.~\eqref{eq:NewEnergy} as a function of the parameters $\alpha_{\mathrm{HO}}$ and $\alpha_{\mathrm{free}}$ for $N=9$ bosons. The left panel is for $\ln(B_2)/N=-2.05$; the right panel is for $\ln(B_2)/N=-2$. Note that both $\alpha$-parameters are defined to be positive.}
    \label{fig:VariationAlphas}
\end{figure}

In this Appendix, we provide additional details on the variational ansatz introduced in Eq.~(\ref{eq:Ansatz}). In particular, we present all relevant equations. For better readability, we use a system of units with $\hbar=m=\omega=1$ here. The explicit form of the variational ansatz is
\\
\begin{equation}
\Psi(\Vec{r})=\frac{1}{\sqrt{\alpha_{\mathrm{HO}}^2+\alpha_{\mathrm{free}}^2+2\alpha_{\mathrm{HO}}\alpha_{\mathrm{free}}\frac{\Bar{C}}{\sqrt{C C_2 }R}}}\left(\alpha_{\mathrm{HO}}\frac{\sqrt{N}}{\sqrt{2\pi C_2}}f_{\mathrm{HO}}(r)+\alpha_{\mathrm{free}}\frac{\sqrt{N}}{\sqrt{2\pi C}R}f_{\mathrm{free}}(r/R)\right)
\end{equation}
\\
\begin{equation*}
    C_2=\int dr r f_{\mathrm{HO}}^2(r),~~~\Bar{C}=\int dr r f_{\mathrm{HO}}(r)f_{\mathrm{free}}(r/R).
\end{equation*}

This ansatz leads to the following expectation value of the Hamiltonian
\\
    \begin{equation}
\label{eq:NewEnergy}
    \begin{split}
    E&=\frac{\alpha_{\mathrm{HO}}^2}{\alpha_{\mathrm{HO}}^2+\alpha_{\mathrm{free}}^2+2\alpha_{\mathrm{HO}}\alpha_{\mathrm{free}}\frac{\Bar{C}}{\sqrt{C C_2 }R}}\left(\frac{N}{2C_2}A_2-\frac{\alpha_{\mathrm{HO}}^2}{\alpha_{\mathrm{HO}}^2+\alpha_{\mathrm{free}}^2+2\alpha_{\mathrm{HO}}\alpha_{\mathrm{free}}\frac{\Bar{C}}{\sqrt{C C_2 }R}}\frac{gN^2B_2}{4\pi C_2}+\frac{1}{2}\frac{N D_2}{C_2}\right)\\
    &+\frac{\alpha_{\mathrm{free}}^2}{\alpha_{\mathrm{HO}}^2+\alpha_{\mathrm{free}}^2+2\alpha_{\mathrm{HO}}\alpha_{\mathrm{free}}\frac{\Bar{C}}{\sqrt{C C_2 }R}}\left(\frac{N}{2C_2R^2}A-\frac{\alpha_{\mathrm{free}}^2}{\alpha_{\mathrm{HO}}^2+\alpha_{\mathrm{free}}^2+2\alpha_{\mathrm{HO}}\alpha_{\mathrm{free}}\frac{\Bar{C}}{\sqrt{C C_2 }R}}\frac{gN^2B}{4\pi C R^2}+\frac{1}{2}\frac{N D R^2}{C}\right)\\
    &+2\frac{\alpha_{\mathrm{HO}}\alpha_{\mathrm{free}}}{\alpha_{\mathrm{HO}}^2+\alpha_{\mathrm{free}}^2+2\alpha_{\mathrm{HO}}\alpha_{\mathrm{free}}\frac{\Bar{C}}{\sqrt{C C_2 }R}}\left( 
\frac{N}{2\sqrt{C C_2}}\Tilde{A}+\frac{1}{2}\frac{ N}{\sqrt{C C_2}}\Tilde{D} \right)\\
&+\frac{\alpha_{\mathrm{HO}}\alpha_{\mathrm{free}}}{\left(\alpha_{\mathrm{HO}}^2+\alpha_{\mathrm{free}}^2+2\alpha_{\mathrm{HO}}\alpha_{\mathrm{free}}\frac{\Bar{C}}{\sqrt{C C_2 }R}\right)^2}\frac{gN^2}{4\pi}\left(4\frac{\alpha_{\mathrm{HO}}^2}{\sqrt{C_2^3C}}\Tilde{B}+4\frac{\alpha_{\mathrm{free}}^2}{\sqrt{C_2C^3}}\Bar{B}+6\frac{\alpha_{\mathrm{free}}\alpha_{\mathrm{HO}}}{\sqrt{C_2^2C^2}}\hat{B} 
 \right)
    \end{split}
\end{equation}
with
\begin{equation*}
    \begin{split}
            A&=\int dr r [f_{\mathrm{HO}}'(r)]^2,~~~B_2=\int dr r f_{\mathrm{HO}}^4(r),~~~D_2=\int dr r^3 f_{\mathrm{HO}}^2(r)\\
            \Tilde{A}&=\int dr r f_{\mathrm{HO}}'(r)f_{\mathrm{free}}'(r/R), ~~~~\Tilde{D}=\int dr r^3 f_{\mathrm{HO}}(r)f_{\mathrm{free}}(r/R)\\
            \Tilde{B}&=\int dr r f_{\mathrm{HO}}^3(r)f_{\mathrm{free}}(r/R), ~~~~\Bar{B}=\int dr r f_{\mathrm{HO}}(r)f_{\mathrm{free}}^3(r/R), ~~~~\hat{B}=\int dr r f_{\mathrm{HO}}^2(r)f_{\mathrm{free}}^2(r/R).
    \end{split}
\end{equation*}

To facilitate the calculation of the $R-$dependent integrals in each step of the minimization calculation, we assume that $R=R_{\mathrm{free}}$ if $R_{\mathrm{free}}<1$ and in all other cases that $R=1$ (see the main text). Next, we minimize the energy, Eq.~\eqref{eq:NewEnergy}, with respect to $\alpha_{\mathrm{HO}}$ and $\alpha_{\mathrm{free}}$ for each interaction strength. We show a plot of the energy as a function of these parameters in Fig.~\ref{fig:VariationAlphas} for two different interaction strengths, for $R_{\mathrm{free}}>1$ (left panel) and for $R_{\mathrm{free}}<1$ (right panel). We can see that the energy is minimized at the boundaries of the plot, i.e., when one of the parameters is zero. For the left panel with $R_{\mathrm{free}}>1$ we can see that it is energetically favorable to have a finite value of $\alpha_{\mathrm{HO}}$ while in the right one $\alpha_{\mathrm{free}}$ is preferred. This implies the sharp transition discussed in the main text; there is no particle number or interaction strength for which both $\alpha$-values are non-zero after the minimization of the energy.

\section{Details on the configuration interaction method}
\label{App:CI}

\begin{figure}[!t]
    \centering
    \begin{minipage}{0.5\linewidth}
    \centering
    \includegraphics[width=1\linewidth]{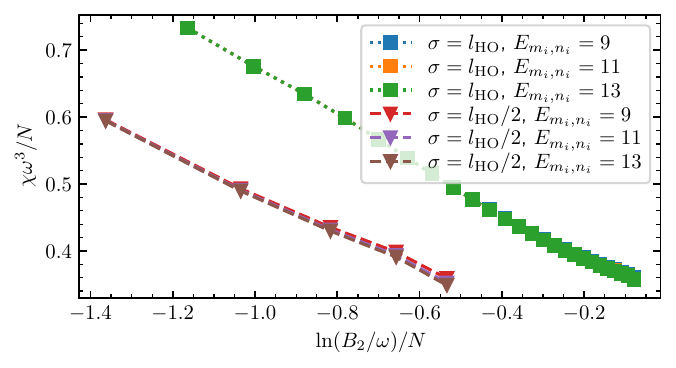}
    \end{minipage}%
    \begin{minipage}{0.5\linewidth}
    \centering
    \includegraphics[width=1\linewidth]{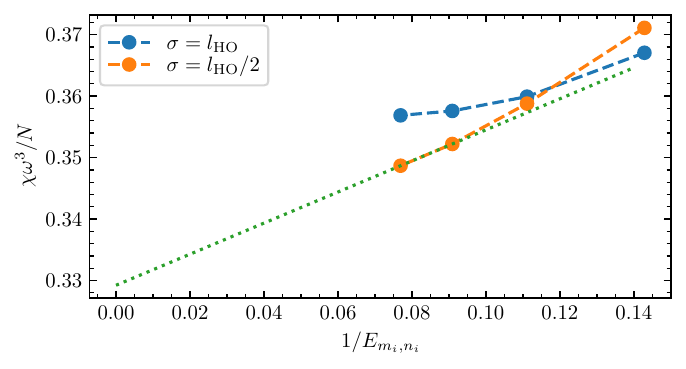}
    \end{minipage}
    \caption{{Convergence pattern for the static response per particle, Eq.~\eqref{eq:Compress}, calculated with a CI method. Left panel: The static response as a function of $\ln(B_2/\omega)/N$ for different one-body cutoffs $E_{m_i, n_i}$. For the interaction, we use a Gaussian interaction (width given in figure legend). Right panel: The static response as a function of $1/E_{m_i, n_i}$ for $N=4$ and the strongest interactions considered in the left panel. The green dotted line corresponds to a linear extrapolation of the data for the two largest cutoffs for $\sigma=l_{\mathrm{HO}}/2$.}}
    \label{fig:APP:ConvergenceCI}
\end{figure}

For the CI calculations, we write the Hamiltonian in the formalism of second quantization:  
\begin{equation}
    H=\sum\limits_{i, j}A_{ij}a_i^\dagger a_j+\sum\limits_{i,j,k,l}V_{ijkl}a_i^\dagger a_j^\dagger a_l a_k,
\end{equation}
with $a_i^\dagger$ ($a_i$) bosonic creation (annihilation) operators. We find the one-body matrix $A_{ij}$ and the two-body matrix $V_{ijkl}$ by expanding them in one-body basis functions. For these functions, we use the eigenfunctions of the non-interacting system, i.e. the eigenfunctions of the 2D harmonic oscillator (for better readability, we use a system of units such that $\hbar=m=\omega=1$):
\begin{equation}
    \Phi_i(\Vec{x})=\frac{1}{\sqrt{2\pi}}e^{im_i\phi}F_{n_i, m_i}(\rho)=\frac{1}{\sqrt{2\pi}}e^{im_i\phi}(-1)^{n_i}\sqrt{\frac{2\Gamma(n_i+1)}{\Gamma(|m_i|+n_i+1)}}e^{-\rho^2/2}\rho^{|m_i|}L_{n_i}^{|m_i|}(\rho^2).
\end{equation}
The index $i$ is the index of the basis state and $m_i$ is the angular component of this index and $n_i$ the radial one. We use polar coordinates with $\rho$ and $\phi$. $L$ is the generalized Laguerre polynomial. Such a state has an eigenenergy of 
\begin{equation}
    E_{m_i, n_i}=2n_i+|m_i|+1\,.
\end{equation}
In the CI-method, we first construct $N-$particle basis states, which are then used to write the Hamiltonian matrix for a given set of parameters. This matrix is diagonalized using the Arnoldi/Lanczos method~\cite{golubmatrix}.

We must truncate the Hamiltonian matrix in numerical calculations. Therefore, we introduce a one-body basis cutoff. We include only one-body basis states whose energy is equal or lower to thirteen. To retrieve the two-body energy in free space, $B_2=-E_2$, we solve the Schrödinger equation in relative coordinates (effectively a one-body problem) with up to 500 basis states.
Furthermore, as our interaction conserves the total momentum, we only include many-body basis states with a total angular momentum of zero. Attractive interactions localize the bosons in the center of the trap, and many basis states are needed to resolve the ground state. To account for this in our calculations, we use a trapping frequency for the one-body basis, which is different from $\omega$: $\omega_{\mathrm{basis}}=4\omega$. This allows us to obtain converged results for stronger interactions. {We show convergence plots for the static response, Eq.~\eqref{eq:Compress}, in Fig.~\ref{fig:APP:ConvergenceCI}. As can be seen, the smaller the Gaussian interaction, the more basis states are needed for convergence. Assuming slow convergence, $\sim 1/E_{m_i, n_i}$, for $\sigma=l_{\mathrm{HO}}/2$ and the strongest considered interaction, we use a linear function to extrapolate the results for the two largest cutoffs to $E_{m_i, n_i}\to\infty$ (the green dotted line). As the assumed convergence behavior is clearly slower than what we observe in practice, see Fig.~\ref{fig:APP:ConvergenceCI}, the extrapolated value allows us to estimate the truncation error to be $\Delta \chi \omega^3/N<0.02$. This implies that the uncertainty in the value of the static response is at a-few-percent level. We conclude that our data is converged at this level of accuracy.}

For a detailed explanation of the employed CI-method we refer to Refs.~\cite{CremonPhDThesis, BjerlinPhdThesis}.

\section{Alternative variational ansatz}
\label{App:OtherAnsatz}

As briefly mentioned in the main text, we also employed a different variational ansatz to study the system. In this approach, both shapes are scaled with the characteristic width $R$:
\begin{equation}
\label{eq:App:2DBosons:AnsatzOther}
    \Psi(\Vec{x})=\mathcal{N}\left(\alpha_{\mathrm{HO}}f_{\mathrm{HO}}(r/R)+\alpha_{\mathrm{free}}f_{\mathrm{free}}(r/R)\right)\,,
\end{equation}
where $\mathcal{N}$ is the normalization coefficient. The advantage of this ansatz, compared to the one used in the main text, is that the energy can be minimized with respect to $\alpha_{\mathrm{HO}}$, $\alpha_{\mathrm{free}}$ and $R$ with similar numerical effort as for the previous ansatz. This eliminates the need to rely on physical intuition to determine $R$. The disadvantage is that it is difficult to justify why both shapes should scale with $R$. We note that although the variational ansatz presented in the main text is physically motivated, the ansatz in this Appendix follows the logic of Ref.~\cite{Hammer2004} more closely. In particular, it allows us to vary the scale parameter $R$.

\begin{figure}[!t]
    \centering
    \begin{minipage}{0.5\linewidth}
    \centering
    \includegraphics[width=1\linewidth]{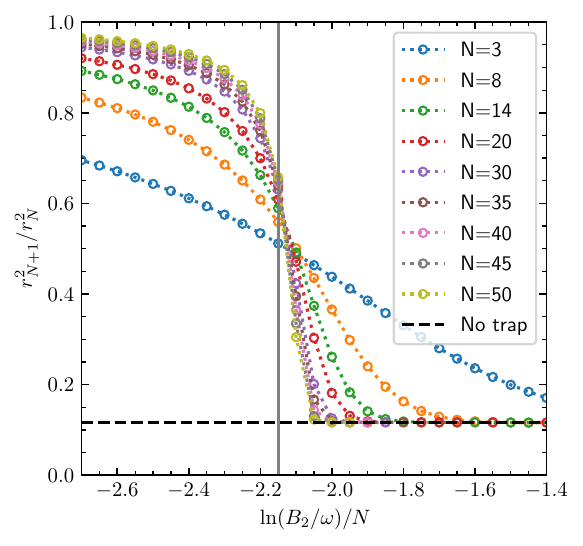}
    \end{minipage}%
    \begin{minipage}{0.5\linewidth}
    \centering
    \includegraphics[width=1\linewidth]{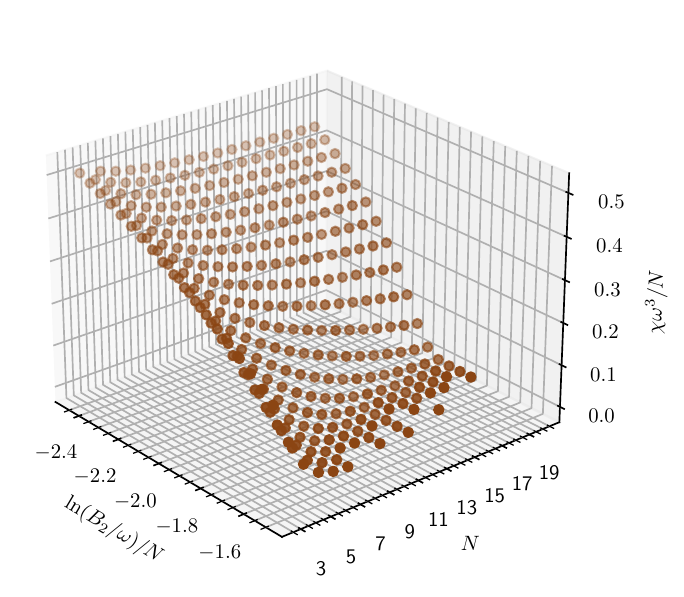}
    \end{minipage}
    \caption{{Ratio of radii (left), $r^2_{N+1}/r^2_{N}$, and static response per particle (right), $\chi/N$, calculated using the ansatz where both shapes are scaled with $R$, Eq.~\eqref{eq:App:2DBosons:AnsatzOther}.
    Symbols are the results obtained from the minimization of the energy. The black dashed line in the left panel is the universal prediction from Ref.~\cite{Hammer2004} and the vertical solid grey line shows the critical interaction strength $\ln(B_2/\omega)/N=-2.148$, Eq.~\eqref{eq:TransitionPoint}. Dotted lines in the left panel are added to guide the eye.}}
    \label{fig:APP:2DBosons:OtherAnsatz}
\end{figure}

The ansatz in Eq.~(\ref{eq:App:2DBosons:AnsatzOther}) leads to a smooth transition between the trap-dominated state and the interaction-dominated one at small particle numbers $N$, which becomes sharper as $N$ is increased. We show this in Fig.~\ref{fig:APP:2DBosons:OtherAnsatz}, where we display the ratio of the mean-square radii. As one can see, in contrast to Fig.~\ref{fig:Radii}, we now observe a smooth crossover between the two states of the system for all particle numbers, with the transition becoming sharper as the particle number increases. We can also see that the critical interaction strength predicted using perturbation theory in the main text, $\ln(B_2/\omega)/N\approx-2.15$, is consistent  with the point where the system transitions from the trap-dominated regime into the interaction-dominated one. 
Additionally, we present the static response in this figure. In this case, the increase in sharpness with $N$ is less pronounced but still visible.

It is beyond the scope of this work to provide a definitive answer as to which ansatz (the one in Eq.~(\ref{eq:Ansatz}) or the one in Eq.~(\ref{eq:App:2DBosons:AnsatzOther})) offers a more accurate description of the system's ground state. However, the variational ansatz presented in this appendix does not exhibit a  discontinuous transition -- even for the largest particle numbers considered -- and thus fails to reproduce the results of Ref.~\cite{tononi2023gastosoliton} for bosons on a sphere. Therefore, we assume that the results based upon Eq.~(\ref{eq:Ansatz}) represent the underlying physics more faithfully. This assumption has to be tested by future research.


In summary, the main conclusions drawn in the main part of our manuscript are supported by the alternative ansatz introduced in Eq.~\eqref{eq:App:2DBosons:AnsatzOther}. However, further research is needed to clarify the nature of the transition between the trap-dominated and interaction-dominated regimes in a harmonic trap.

\bibliography{refs}

\end{document}